\documentclass[twocolumn,showpacs,preprintnumbers,amsmath,amssymb]{revtex4}
 
 \usepackage{fancyheadings}
 
 \usepackage{amsmath} 
\usepackage{amsfonts} 
\usepackage{amssymb}
\usepackage{graphicx} 

\begin{document}

\title{Classical Time Crystals}
\author{Alfred Shapere$^1$ and Frank Wilczek$^2$\vspace*{.1in}}
\affiliation{$^1$Department of Physics and Astronomy,
University of Kentucky, Lexington, Kentucky 40502 USA\\
$^2$Center for Theoretical Physics,
Department of Physics, Massachusetts Institute of Technology,
Cambridge, Massachusetts 02139 USA}
\vspace*{.3in}

\begin{abstract}
We consider the possibility that classical dynamical systems display motion in their lowest energy state, forming a time analogue of crystalline spatial order.   Challenges facing that idea are identified and overcome.  We display arbitrary orbits of an angular variable as lowest-energy trajectories for nonsingular Lagrangian systems.   Dynamics within orbits of broken symmetry provide a natural arena for formation of time crystals.  We exhibit models of that kind, including a model with traveling density waves.   
\end{abstract}



\maketitle

\thispagestyle{fancy}


In this paper we will investigate a cluster of issues around the question of whether time-independent, conservative classical  systems might exhibit motion in their lowest energy states.    Fully quantum systems are the subject of a companion paper \cite{wilczek}.    Related issues have been raised in a cosmological context \cite{ArmendarizPicon:1999rj}\cite{a-hetal}, but those investigations consider quite different aspects, in which the time dependence introduced by the expansion of the universe plays a significant role.  (The term ``time crystal'' has been used previously to describe periodic phenomena in other contexts \cite{winfree, srivastava-widom}.) 
 
{\it General considerations.} 
When a physical solution of a set of equations displays less symmetry than the equations themselves, we say the symmetry is spontaneously broken by that solution.   Here the meaning of ``physical solution'' can be interpreted differently in different contexts, but one interesting case, that will concern us here, is of the lowest energy solutions of a time-independent, conservative, classical dynamical system.   If such a  solution exhibits motion, we will have broken time translation symmetry spontaneously.   If the dynamical variable is an angular variable, then the motion will be periodic in time, so the time-translation symmetry is not entirely lost, but only reduced to a discrete subgroup.   Spatial periodicity is, of course, associated with formation of ordinary crystals, so it is natural and suggestive to refer to the formation of time crystals.   

It is very easy to construct simple Lagrangians or Hamiltonians whose lowest energy state is a spatial crystal.  With $\phi(x)$ an angular variable, the potential energy functions
\begin{eqnarray}\label{crystalPotentials}
V_1(\phi) ~&=&~ - \kappa_1 \frac{d\phi}{dx} + \frac{\lambda_1}{2} (\frac{d\phi}{dx})^2 \nonumber \\
V_2(\phi) ~&=&~ - \frac{\kappa_2}{2}( \frac{d\phi}{dx})^2 + \frac{\lambda_2}{4}( \frac{d\phi}{dx})^4
\end{eqnarray}
with all the Greek coefficients positive, are minimized for $\frac{d\phi_1}{dx} = \frac{\kappa_1}{\lambda_1}, \frac{d\phi_2}{dx}= \pm \sqrt{\frac{\kappa_2}{\lambda_2}}$ respectively.  In both cases the spatial translation symmetry of the original potential is spontaneously broken; in the second case inversion symmetry is broken as well.  The combined inversion $\phi(x) \rightarrow - \phi(-x)$ is preserved in both cases, as is a combined internal space-real space translation $\phi (x) \rightarrow \phi (x + \epsilon) - \frac{d\phi}{dx} \epsilon$.

From this one might surmise that time crystals are likewise easy to construct, at least mathematically.  Moreover, higher powers of velocities
appear quite naturally in models that portray the effects of finite response times, as we replace 
\begin{equation}
\bigl( (\phi (t) - \phi (t - \delta ) \bigr)^n ~\tilde{\rightarrow}~  \delta^n {\dot \phi}^n
\end{equation}
On second thought, however, reasons for doubt appear.   Speaking broadly, what we're looking for seems perilously close to perpetual motion.   Also, if the dynamical equations conserve energy, then the existence of a minimum-energy solution where the variables trace out an orbit implies that the energy function assumes its minimum value on a whole curve in $(\phi, \dot \phi)$ space -- not, as we expect generically, at an isolated point.    

{\it Dynamical equations.} That easy/impossible dichotomy carries over into the dynamical equations.   If one simply turns the space derivatives in Eqn. (\ref{crystalPotentials}) into time derivatives, then the resulting Lagrangians
\begin{eqnarray}\label{lagrangians}
L_1 (\phi,\dot\phi) ~&=&~ - \kappa_1 \dot \phi  + \frac{\lambda_1}{2} \dot \phi^2 \nonumber \\
L_2(\phi,\dot\phi) ~&=&~ - \frac{\kappa_2}{2}\dot \phi^2 + \frac{\lambda_2}{4}\dot \phi^4
\end{eqnarray}
are associated with the energy functions
\begin{eqnarray}\label{energies}
E_1 (\phi,\dot\phi) ~&=&~  \frac{\lambda_1}{2} \dot \phi^2 \nonumber \\
E_2(\phi,\dot\phi) ~&=&~ - \frac{\kappa_2}{2}\dot \phi^2 + \frac{3 \lambda_2}{4}\dot \phi^4 \, .
\end{eqnarray}
The first of these is minimized at $\dot \phi_1 = 0$, the second at $\dot \phi_2 = \pm \sqrt {\frac{\kappa_2}{3\lambda_2}}$.   So the analogue of our first symmetry-breaking example in Eqn.(\ref{crystalPotentials})  has collapsed, but the second survives, with a quantitative change.   

On the other hand if we convert the space derivatives in Eqn. (\ref{crystalPotentials}) into momenta, the resulting Hamiltonians are
\begin{eqnarray}
H_1(p,\phi) ~&=&~ - \kappa_1 p   + \frac{\lambda_1}{2} p^2 \nonumber \\
H_2(p,\phi) ~&=&~ - \frac{\kappa_2}{2} p^2  + \frac{\lambda_2}{4} p^4 \, . 
\end{eqnarray}
We find precisely the original algebraic structure for the minimum-energy solutions, {\it viz}.  $p_1 = \frac{\kappa_1}{\lambda_1}, p_2 = \pm \sqrt{\frac{\kappa_2}{\lambda_2}}$ respectively.    Their physical implications are entirely different, though.  Indeed, they correspond to $\dot \phi_1 = \dot \phi_2 = 0$: thus no symmetry breaking occurs, in either case.     

This disappointing consequence of the Hamiltonian formalism is quite general.   Hamilton's equations of motion
\begin{eqnarray}\label{hamilton}
\dot p_j ~&=&~ -\frac{\partial H}{\partial q^j} \nonumber \\
\dot q^j ~&=&~ \ \ \ \frac{\partial H}{\partial p_j}
\end{eqnarray}
indicate that the energy function $E(p_j (0), q^j (0)) = H(p_j (0), q^j (0))$, regarded as a function of the dynamical variables at a chosen initial time, is minimized for trajectory with $\dot p_j = \dot q^j =0$, since the gradients on the right-hand side of Eqn. (\ref{hamilton}) vanish.  

How do we reconcile this very general null result in the Hamiltonian approach, with our positive result in the Lagrangian approach?  The point is that the Lagrangian $L_2$, which gave symmetry breaking, cannot be converted into a Hamiltonian smoothly.   Indeed, expressing the algebraic recipe for the Hamiltonian
\begin{equation}
H(p,\phi) ~=~ p \dot \phi - L ~=~ p \dot \phi + \hbox{$\frac{\kappa}{2}$} \dot \phi^2 - \hbox{$\frac{1}{4}$} \dot \phi^4
\end{equation}
(in which we have set $\lambda_2 =1$ for simplicity and dropped all `2' subscripts) 
as a function of 
\begin{equation}
p ~=~ \frac{\partial L}{\partial \dot \phi} ~=~ \dot \phi^3 - \kappa \dot \phi
\end{equation}
leads to a multi-valued function \cite{HTZ}, with cusps where $\frac{\partial p}{\partial \dot \phi} =0$, {\it i.e.} $p = \mp \frac{2\kappa^{3/2}}{3^{3/2}}$, corresponding precisely to the energy minima $\dot \phi = \pm \sqrt{{\kappa}/{3}}$.  (See Figure 1.)   For $\kappa \leq 0$, $ H(p)$ is regular, but as $\kappa$ passes through zero there is a swallowtail catastrophe.   

\begin{figure}[ht]
\begin{center}
\includegraphics[scale=0.6]{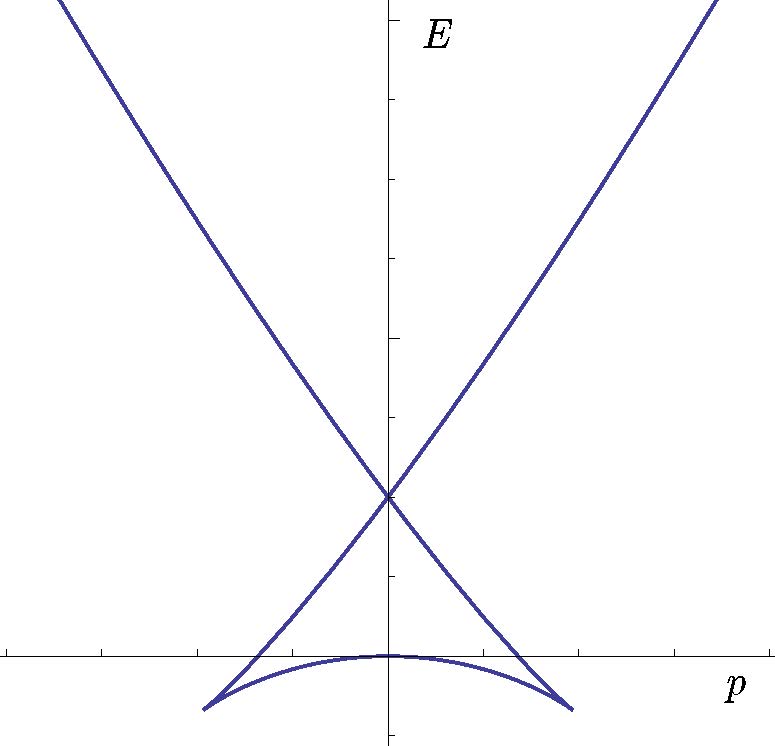}
\caption{Energy is a multivalued function of momentum.}
\end{center}
\end{figure}

At the cusps the usual condition that the gradient should vanish at a minimum does not apply, and so our null result for smooth Hamiltonian systems is avoided.

For classical physics the Lagrangian formalism is adequate, so let us follow that direction out further.   A logical next step would be to add a potential $V(\phi)$ to $L$.    Doing that, however, leads us directly into the problem with energy conservation that we anticipated earlier.   Minimizing $V$, we will find a preferred value for $\phi = \phi_0$, but minimizing the kinetic part will favor motion in $\phi$, and there is a conflict.   

We can elucidate this issue as it arises for a general Lagrangian system.  Suppose that 
the energy function of  a system with many degrees of freedom is minimized by nonzero velocities $\dot\phi_0^k\ne 0$, so that
\begin{eqnarray}\label{dE}
0 ~&=&~ \left.\frac{\partial E}{\partial \dot \phi^k}\right|_{\dot\phi^k_0} 
~=~ \left.\left(\frac{ \partial^2 L} {\partial \dot \phi^k \partial \dot \phi^j } \right)\right| _{\dot \phi^k_0} \dot \phi^j_0 ~.
\end{eqnarray}
Then in the equations of motion 
\begin{equation}\label{eom}
0 ~=~ \frac{d}{dt} ( \frac{\partial L}{\partial \dot \phi^k} ) - \frac{\partial L}{\partial \phi^k} ~=~ 
\bigl(\frac{\partial^2 L}{\partial \dot \phi^j \partial \dot \phi^k} \, \bigr) \ddot \phi^j  + \dots
\end{equation}
the coefficient of the acceleration  in the direction $\ddot \phi^j \propto \dot \phi^j_0 $ vanishes at $\dot {\phi}^k_0$.
In that case the equations of motion, which generally serve to determine the accelerations, require supplementation.   (As we shall discuss below, there are physically interesting models that avoid any singularities of this type.) 

{\it Brick Wall Solutions}: Upon integrating
\begin{equation}\label{energyV}
E~=~\hbox{$\frac34$} \dot\phi^4 - \hbox{$\frac{\kappa}2$} \dot\phi^2+V(\phi)
\end{equation}
directly we obtain
\begin{equation}\label{tphi}
t(\phi) ~=~ \int^\phi \frac{d\phi}{\pm\sqrt{\frac{\kappa}{3}\pm \sqrt{\left(\frac{\kappa}{3}\right)^2+\frac43(E-V(\phi))}}}
\end{equation}
where the $\pm$ signs are independent.   

The argument of the inner square root is non-negative if and only if 
$V(\phi)\le E+{\kappa^2}/{12} \equiv \Delta $,
where $\Delta \equiv E - E_0 \ge 0$ is the energy above the minimum kinetic energy $E_0= -\frac{\kappa^2}{12}$.   The inequality is saturated when $\dot\phi = \pm\sqrt{\frac{\kappa}{3}}$, i.e., when the kinetic energy is minimized.  Close to a point $\phi_{\rm t}$ where this happens,  
\begin{equation}
\dot\phi ~\approx~ \pm\sqrt{\hbox{$\frac{\kappa}{3}$}}\pm\sqrt{\hbox{$\frac1\kappa$} V'(\phi_{\rm t})(\phi_{\rm t} - \phi)} \, . 
\end{equation}
Since $\phi$ cannot continue past $\phi_{\rm t}$ without violating the bound $V(\phi)\le  \Delta $, it suddenly reverses direction, $\dot\phi=\pm \sqrt{\frac{\kappa}{3}}\to  \mp\sqrt{\frac{\kappa}{3}}$.  Such a reversal conserves energy, but requires a sudden jump in  momentum.    This is analogous to the turning point of a ``brick-wall" potential enforced by an infinitely massive source.  Unless $\phi_{\rm t}$ is an extremum of $V(\phi)$, the acceleration diverges at $\phi_{\rm t}$, as required by the equations of motion (\ref{eom}).

Small oscillations about the minimum of a generic potential $V(\phi)\approx \frac12 \mu(\phi-\phi_0)^2$ 
exhibit turning points of this type, with bounded orbits that oscillate between $\phi_{\rm t}= \phi_0- \sqrt{ 2\Delta /\mu}$ and $\phi_0+ \sqrt{ 2\Delta /\mu}$ .   In the limit of small $\Delta$, the orbits ricochet about the minimum, with nearly constant speed $|\dot\phi|=\sqrt{\frac{\kappa}{3}}$, reconciling the apparently contradictory conditions $\dot\phi= \pm\sqrt{\frac{\kappa}{3}}$ and $\phi= \phi_0$.

More conventional turning points arise if the inner $\pm$ sign in (\ref{tphi}) is negative, i.e. for $V(\phi)\ge E$.  Now $\dot\phi_{\rm t}=0$ at turning points where $V(\phi_{\rm t}) = E$, and the particle changes direction smoothly.   


{\it {$fgh$} model}: Lagrangians of the form 
\begin{equation}\label{fghLagrangian}
L ~=~ f \dot \phi^4 + g \dot \phi^2 + h
\end{equation}
for functions $f(\phi), g(\phi), h(\phi)$ lead to energies of the form
\begin{eqnarray}\label{fghEnergy}
E ~&=&~ 3 f\dot \phi^4 + g \dot \phi^2 - h \nonumber \\
~&=&~ 3f ( \dot \phi^2 + \frac{g}{6f})^2 - \frac{g^2}{12f} - h 
\end{eqnarray}
(Note that $f$ may be absorbed into a redefinition of $\phi$; then for constant $g$ this reduces to the model of Eqn.\ (\ref{energyV}).)
If
\begin{equation}\label{tuning}
\frac{g^2}{12f} + h ~=~ {\rm const.}
\end{equation}
the energy will be minimized along the curve $\dot \phi^2 + \frac{g}{6f} \ = \ 0$, for any $f>0, g<0$.   Thus we have solutions 
\begin{equation}
\dot \phi ~=~ \pm \sqrt{-\frac{g}{6f}}
\end{equation}
This construction demonstrates that any orbit with a velocity that does not change sign can be realized, in many ways, as the stable minimum energy solution to an appropriate, reasonably simple Lagrangian.   

Choosing the constant in (\ref{tuning}) to be zero, constancy of the energy $E \geq 0 $ leads to 
\begin{equation}\label{fghDynamics}
\dot \phi^2 +  \frac{g}{6f} ~=~ \pm \sqrt{\frac{E}{f}} 
\end{equation}
This equation is of a familiar form; it expresses the conservation of a pseudo-energy $\tilde E$ for a particle with mass $m=\frac{1}{2}$ and a two-branched $E$-dependent potential, according to
\begin{eqnarray}
\tilde E ~&=&~ \dot \phi^2 + V(E, \phi) \label{pseudoEnergy} \\
V(E, \phi) ~&=&~ \frac{g}{6f} \mp \sqrt{\frac{E}{f}} \label{effectiveV}
\end{eqnarray}
This result allows us to infer the qualitative dynamics, based on familiar mechanical concepts.  Perhaps the most interesting question is the existence, or not, of turning points.   Putting $\dot \phi = 0$ into Eqn. (\ref{pseudoEnergy}) we find that $E = - g^2 (\phi_{\rm t})/6 f (\phi_{\rm t}) \equiv V_{\rm max}$ evaluated at the turning point(s) $\phi_{\rm t}$.   The motion is confined to a region where $V \leq V_{\rm max}$. Thus the model can support motions in which the velocity changes sign, but these motions require {\it higher\/} energy that the minimal orbit, which is unidirectional.   Actually nothing in our analysis of this model has depended on the periodicity of $\phi$; upon dropping that assumption, we find the curious situation that some unbounded motions have smaller energy than any bounded motion.  

{\it Avoiding Singularities}:  If we relax the condition Eqn. (\ref{tuning}), by allowing a non-constant $W \equiv - \frac{g^2}{12f} - h$, we find that there are initial conditions for which the equations of motion eventually become singular, as discussed previously.   We are guaranteed to avoid such conditions if in Eqn. (\ref{fghEnergy}) we require
\begin{equation}\label{safeEnergy}
E \geq \ W_{\rm max}
\end{equation}
(Note that in any case $E \geq  W_{\rm min}$.)     Thus we have models that work smoothly for high energy, but become singular at low energy.  That is the opposite of the usual philosophy of effective field theory; however it does correspond to the use of perturbative QCD.   

We might also expect that the quantum-mechanical versions of these models might be more robust, in that the uncertainty in position and velocity might smooth over a small region of singularity.    Genuine quantization of such models -- or, for that matter, of the tuned $fgh$ models and the natural, locked models to come -- presents interesting issues.   The Lagrangian formulation is adequate for path integral quantization, and the higher derivative terms tend to damp the contribution of the most irregular paths, particularly if we continue the time to have a small negative imaginary part.   So there are no obvious show-stoppers, but no existence proofs either.   The Hamiltonian formulation poses different issues.  As we've seen, in interesting cases the Hamiltonian is a multi-valued function of the momentum.   This implies that the momentum does not provide a complete set of commuting observables.  Wave functions must be defined as functions of expanded spaces.   We will report work on this subject elsewhere.         

What we can discuss simply is semiclassical quantization.   Thus we consider orbits obeying a Bohr-Sommerfeld condition
\begin{equation}
S ~=~ \oint p \,d\phi ~=~ \int_0^{2\pi} (\dot \phi^3 - \kappa \dot \phi)d\phi ~=~ 2\pi\hbar (n + \delta)
\end{equation}
with $n$ an integer, and $\delta$ a correction for turning points.   (For simplicity we specialize here to $f=\frac{1}{4}$, $g = -\frac{\kappa}{2}$, and $h = 0$.)   If we ignore, at first, the potential, then the minimal energy orbit is at $\dot \phi = \pm \sqrt{\frac{\kappa}{3}}$, and for it   $$S ~=~ \mp 2 \pi \frac{2\kappa^{3/2}}{3^{3/2}}.$$   
If this expression is not equal to $2\pi \hbar (n+\delta)$, the quantization condition will lead us to a nearby higher-energy orbit for the ground state, with some $| n | \gg 1$ in the relevant (semiclassical) limit.    If the potential is small enough, that extra energy will be enough to enforce Eqn.~(\ref{safeEnergy}), and keep us out of the region where the equation of motion breaks down.    Wave packets constructed from $n$ near the preferred value will describe, approximately, the motion prescribed by the classical dynamics.


{\it Naturally Flat Directions; Double Sombrero}:  It can be natural to have energy constant along an orbit, if the points of the orbit are related by symmetry.   If we want this situation to occur along a trajectory representing non-trivial motion in the minimum-energy state, then the points assumed at different times must be related by symmetry transformations, which implies that none of them is invariant.  So we will be looking at models with spontaneously broken symmetry.  

Consider first a Lagrangian with a ``sombrero'' kinetic term:
\begin{align}\label{sombrero}
L &=  \hbox{$\frac14$} ({\dot \psi}_1^2+{\dot \psi}_2^2 - \kappa)^2-V(\psi_1,\psi_2)
\end{align}
The matrix of second derivatives of $L$ with respect to $\dot\psi_i$ appearing in Eqn.\ (\ref{dE}) is 
\begin{equation}
\frac{\delta^2 L}{\delta \dot\psi_i \delta\dot\psi_j}= 
\left(\begin{array}{cc}
3\dot\psi_1^2+\dot\psi_2^2-\kappa&2\dot\psi_1\dot\psi_2\\
2\dot\psi_1\dot\psi_2 &\dot\psi_1^2+3\dot\psi_2^2-\kappa
        \end{array}\right)
\end{equation}
This has a zero eigenvalue with eigenvector $\begin{pmatrix}\dot\psi_1\\ \dot\psi_2\end{pmatrix}$ precisely when  $v^2\equiv\dot\psi_1^2+\dot\psi_2^2 = \kappa/3$. 

If the potential $V$ has a one-parameter family of degenerate minima,  the minimum-energy solution will move along the trough of $V$ at constant speed $\sqrt{\kappa/3}$.  The potential 
\begin{equation}
V=-\frac{\mu}2 (\psi_1^2+\psi_2^2)+\frac{\lambda}4 (\psi_1^2+\psi_2^2)^2 \
\end{equation}
is symmetric under $\psi_1$-$\psi_2$ rotations and, combined with the kinetic term in Eqn.~(\ref{sombrero}),
defines a ``double sombrero" model, with circular motion at constant speed in the lowest-energy state.

Alternatively we may rewrite this model and its generalizations in terms of  polar fields $\rho$ and $\phi$, where $ \psi_1+i\psi_2 = \rho e^{i\phi}\equiv \varphi.$
Then the double sombrero Lagrangian takes the form
\begin{equation}
L=  \hbox{$\frac14$} (\dot\rho^2+\rho^2\dot\phi^2 - \kappa)^2+\hbox{$\frac\mu{2}$}\rho^2-\hbox{$\frac{\lambda}4$}  \rho^4 \, . 
\end{equation}
If $\rho$ is set equal to its value $\sqrt{2\mu/\lambda}$ at the minimum of $V(\rho)$, this reduces to our original Lagrangian (\ref{lagrangians}). 
Generalizing, any Lagrangian with a kinetic term that is a polynomial in $\dot\phi$, $\dot\rho$, and $\rho$, and a potential energy depending only on $\rho$, will preserve the symmetry $\phi\to \phi+\eta$.  

{\it Charge and Locking}: 
The charge operator associated with the original (broken) symmetry is $Q = -\int i (\varphi^* \pi_{\varphi^*} - \varphi \pi_\varphi) 
$ where $\pi_\varphi = \frac{\partial L}{\partial \dot\varphi}$ depends only on $\dot\phi$ and $\rho$.  Thus in states with constant, non-vanishing values of $\rho$ and $\dot \phi$ we have a non-zero, uniform density of $Q$.    This is significant in two ways:   

First: If we suppose that our system is embedded in a larger symmetry-conserving bath and undergoes a transition to the symmetry-breaking state, {\it e.g}. that it is a material body cooled through a phase transition, then the transition will necessarily be accompanied by radiation of an appropriate balancing charge.  

Second: Although invariance under both infinitesimal time-translation $\phi (t ) \rightarrow \phi (t+\epsilon )$ and infinitesimal phase (charge) translation $\phi \rightarrow \phi + \eta$ are broken by constant-$\dot \phi$ solutions 
$\phi (t) = \omega t + \beta$,  the  combined transformation with $\omega \epsilon + \eta = 0$ leaves the solution invariant.  Thus there is a residual ``locked'' symmetry.    To exploit it, we can go to a sort of rotating frame, by using the shifted Hamiltonian
$\widetilde H ~=~ H - {\omega} Q$
to compute the evolution \cite{a-hetal, Nicolis}.   (Here we normalize $Q$ so that $\varphi$ has unit charge.)  In the rotating frame, the equations of motion will not contain any explicit time dependence, but there will be a sort of effective chemical potential (associated however with a {\it broken\/} symmetry).   
The most interesting effects will arise at interfaces between the locked phase and the normal phase, or between different locked phases, as exemplified in the preceding paragraph.  

{\it Space-Time Structure; More Complex States}:
We can also contemplate slightly more complex examples, that support qualitatively different, richer physical effects.   If there is a potential for $\nabla \varphi$, or ultimately for $\nabla \rho$, that favors gradients, then we can have a competition between the energetic desirability of putting $\rho$ at the energetic minimum and accommodating non-zero gradients.   Unlike the case of time derivatives, there is no general barrier to reaching a stable compromise.    To keep things simple, let us suppress the underlying $\varphi$ structure  and consider the potential
\begin{equation}\label{withgradient}
V(\rho) ~=~ \frac{\kappa_1}{2} \left(1 - a \rho^2 -  b \left(\frac{d\rho}{dx}\right)^2\right)^2
\end{equation}
with $a, b>0$.  This potential is minimized by
$$\rho_0(x) = \sqrt{\frac{1}{a}} \, \sin (\sqrt{\frac{a}{b}} x + \alpha),$$
which  reduces the translation symmetry to a discrete subgroup.   Constant $\dot \phi$ produces a charge density wave. 

If we also have a term of the form
\begin{equation}\label{gradientLocking}
V_{\rm gradient} ~=~ \frac{\kappa_2}{2} \left(\frac{d\phi}{dx} - \mu \frac{d\rho}{dx}\right)^2
\end{equation}
then at the minimum $\phi_0(x)$ will develop spatial structure as well, according to $\phi_0(x) = \mu \rho_0(x) + \beta$, breaking the phase (charge) symmetry completely.  (Note that $V_{\rm gradient}$ respects the symmetry $\phi \rightarrow \phi + \eta$.)

We can engineer similar phenomena involving $\dot \phi$ most easily if we work at the level of the energy function.  One can derive general energy functions involving powers of $\dot \phi$ from Lagrangians of the same kind, so long as there are no terms linear in $\dot \phi$.   Thus if we have additional term 
\begin{equation}\label{kineticLocking}
E_{\rm kinetic}(\phi) ~=~ \frac{\kappa_3}{2} \left(\left(\frac{d\phi}{dx}\right)^2 - \frac{1}{v^2} \dot \phi^2\right)^2 
\end{equation}
then at the minimum we have 
\begin{eqnarray}\label{phi0}
\phi_0 (x, t) ~&=&~  \mu \rho_0(x , t ) + \beta  \label{phi0} \\
\rho_0 (x, t) ~&=&~  \sqrt{\frac{1}{a}} \, \sin \Bigl(\sqrt{\frac{a}{b}} (x \pm vt ) + \tilde \alpha \Bigr) .  \label{modifiedRho0} 
\end{eqnarray}
Here in Eqn. (\ref{modifiedRho0}) we have adapted our solution $\rho_0(x)$ for the potential 
(\ref{withgradient})  by taking $\alpha = \pm vt + \tilde \alpha$.  In doing this we assume that the energy intrinsically associated with time derivatives of $\rho$ vanishes (or that it is dominated by the locking effects of 
Eqns.~ (\ref{gradientLocking}, \ref{kineticLocking})).   Both spatial and time translation are spontaneously broken, as is reflected in the disposable constants $\tilde \alpha, \beta$, and so is time-reversal $T$, as reflected in the disposable sign.   

Combining Eqns.~(\ref{phi0}, \ref{modifiedRho0}), we now have a traveling charge density wave.  Thus this example exhibits its time-dependence in a physically tangible form.   The residual continuous symmetry is reduced to a combined discrete time-space-charge transformation.  Although our construction has been specific and opportunistic, it serves to establish the existence of a universality class that, since it is characterized by symmetry, should be robust.   It is noteworthy that cyclic motion of $\phi$ in internal space has given rise to linear motion in physical space.   

{\it Relativistic Lagrangians}: All of our constructions above have been nonrelativistic.   In a relativistic theory there are relations among the coefficients of time and space gradient terms.   The relativistic quartic term $L \propto ( (\partial_0 \phi)^2 - (\nabla \phi)^2 )^2$ leads to an energy that is unbounded below, for large gradients of one kind or another.  But use of a sextic enables positive energy.   Indeed, the energy function for $( (\partial_0 \phi)^2 - (\nabla \phi)^2 )^n$ is 
\begin{equation}( (2n-1) (\partial_0 \phi )^2 + (\nabla \phi)^2 )  ( (\partial_0 \phi)^2 - (\nabla \phi)^2 )^{n-1}.
\end{equation}  
\\
For $n$ odd this is semi-positive definite, with a zero at $(\partial_0 \phi)^2 = (\nabla \phi)^2 $ unless $n = 1$.   For $n$ even it has no definite sign.  Bounded energy requires only that the leading term have odd $n$ and a positive coefficient and that the coefficient of the $n=1$ term be non-negative.   
This consideration seems to have been overlooked and might help to constrain the models of \cite{ArmendarizPicon:1999rj} \cite{a-hetal}.

{\it Acknowledgements}:  We thank Maulik Parikh for helpful discussions.   AS is supported in part by NSF Grants PHY-0970069 and PHY-0855614.  FW is supported in part by DOE grant DE-FG02-05ER41360.

\end{document}